\documentclass[12pt]{iopart}
\pdfoutput=1
\usepackage{iopams,amsthm,hyperref,datetime,color,bbm}

\expandafter\let\csname equation*\endcsname\relax
\expandafter\let\csname endequation*\endcsname\relax
\usepackage{amsmath}

\newdateformat{mydate}{\THEDAY\  \monthname[\THEMONTH] \THEYEAR}
\theoremstyle{plain}
\newtheorem*{thm}{Theorem}

\begin{document} 

\title[Local and entanglement entropy of the free Fermi gas]
{Large-scale behaviour of local and entanglement 
entropy of the free Fermi gas at any temperature\footnote{Published in slightly different form as \emph{ J.\,Phys.\,A: Math. Theor. {\bf 49} (2016) 30LT04 (9pp)}. \newline [There it should read $(T_{0}/T)^{1/2}$ in Eq.~(14b).] See also arXiv: 1501.03412v3 [quant-ph].}} 

\author{Hajo Leschke$^{1,2}$, Alexander V.~Sobolev$^3$ and \mbox{Wolfgang Spitzer$^2$}}

\address{$^1$ Institut f\"ur Theoretische Physik, 
Universit\"at Erlangen--N\"urnberg, Staudtstra\ss e~7, 91058~Erlangen, Germany}
\address{$^2$ Fakult\"at f\"ur Mathematik und Informatik, 
FernUniversit\"at Hagen, Universit\"atsstra\ss e~1, 58097~Hagen, Germany}
\address{$^3$ Department of Mathematics, 
University College London, Gower Street, London, WC1E 6BT, United~Kingdom}

\ead{\mailto{hajo.leschke@physik.uni-erlangen.de}, 
\mailto{a.sobolev@ucl.ac.uk}, 
\mailto{wolfgang.spitzer@fernuni-hagen.de}}

\begin{abstract} The leading asymptotic large-scale behaviour of the \emph{spatially bipartite entanglement entropy} (EE) of the free Fermi gas infinitely extended in multidimensional Euclidean space at zero absolute temperature, $T=0$, is by now well understood. Here, we present and discuss the first rigorous results for the corresponding EE of thermal equilibrium states at $T>0$. The leading large-scale term of this \emph{thermal} EE turns out to be twice the first-order finite-size correction to the infinite-volume thermal entropy (density). Not surprisingly, this correction is just the thermal entropy on the interface of the bipartition. However, it is given by a rather complicated integral derived from a semiclassical trace formula for a certain operator on the underlying one-particle Hilbert space. But in the zero-temperature limit \mbox{$T\downarrow0$}, the leading large-scale term of the thermal EE considerably simplifies and displays a $\ln(1/T)$-singularity which one may identify with the known logarithmic enhancement at $T=0$ of the so-called area-law scaling.
\end{abstract}

{}

\hfill{\footnotesize Date of this version: 27 June 2016}

\noindent
{\footnotesize \it
In memory of Enrico Fermi (1901--1954) \\ on the occasion of the 90$^{\text{th}}$ birthday of the ideal Fermi gas}

\section{Introduction and main results.} In recent years, entanglement entropy has turned out to be a useful and much studied quantifier of nonclassical correlations between subsystems in composite quantum systems \cite{HHHH}. In particular, given the (pure) ground state of a spatially large many-particle system and reducing (or localising) it to a spatial subregion $\Omega$, we denote the von Neumann entropy of the resulting (mixed) substate by $S(0,\Omega)$ and call it the local ground-state entropy. The \emph{spatially bipartite entanglement entropy} (EE), defined for a \emph {bounded} $\Omega$ as the (quantum) mutual information relative to the complement of $\Omega$ and denoted by $H(0,\Omega)$, then simply equals $2S(0,\Omega)$ by the purity of the ground state. This (ground-state) EE quantifies, to some extent, how strongly all particles within $\Omega$ are correlated with all those outside $\Omega$. It is thus, for example, well suited to detect  long-range correlations near the critical point of a quantum phase transition by enlarging $\Omega$, see \cite{Sach,Vidal,Chak,Pastur}.

For a many-particle system without long-range interactions the ground-state EE 
\begin{equation}\label{1}
H(0,L\Omega) = 2S(0,L\Omega)\,,\quad L\geq 1
\end{equation} 
is widely believed~\cite{ECP,H} to grow to leading order proportional to the area $|\partial\Omega| L^{d-1}$ of the boundary surface $\partial(L\Omega)$ of the scaled region $L\Omega$ as the (dimensionless) scaling parameter $L$ tends to infinity, $L\to\infty$. Here, $d=1,2,3,\ldots$ is the spatial dimension of $\Omega$. If the particles are fermions and if there is no spectral gap above their ground-state energy in the infinite-volume limit, the effective long-range correlations lurking in the \mbox{(Pauli--)}Fermi--Dirac statistics are expected to slightly enhance the large-scale growth of $H(0,L\Omega)$ by a logarithmic factor $\ln(L)$. Indeed, for the free Fermi gas infinitely extended in $d$-dimensional Euclidean space $\mathbb R^d$ such a large-scale growth of the ground-state EE with a precise and rather explicit prefactor has been proved with full mathematical rigour \cite{LSS1}, thereby confirming a stimulating conjecture by Gioev and Klich \cite{GiKl}. 

In this note, we present and discuss the first rigorous results on the EE of the free Fermi gas in $\mathbb R^d$ in the state of thermal equilibrium at  nonzero temperature, $T>0$, and chemical potential $\mu\in\mathbb R:=\mathbb R^1$. The latter we mostly suppress for notational simplicity, but also because we often consider thermal quantities for fixed (mean) particle-number density $\rho>0$. In contrast with the ground-state or $T=0$ case, the EE at $T>0$, denoted by $H(T,\Omega)$, must not be expected to be just twice the local thermal entropy $S(T,\Omega)<\infty$, since thermal states are mixed ones. However, in trying to define $H(T,\Omega)$ as the mutual information 
\begin{equation} \label{2}
H(T,\Omega) = S(T,\Omega) + ``S(T,\mathbb R^d\setminus\Omega) - S(T,\mathbb R^d)"
\end{equation}
of the bipartition, we are confronted with the  two infinities  $S(T,\mathbb R^d\setminus\Omega) = S(T,\mathbb R^d) = \infty$ due to the additivity of macroscopic thermal entropy. We solve this problem by rewriting the right-hand side of \eqref{2} in a mathematically and physically reasonable way as a sum of two finite (that is, not infinite) differences, see \eqref{def:Delta} and \eqref{def:EE} below. By construction, $H(T,L\Omega)$ is then well-defined and exhibits a leading term proportional to $L^{d-1}$ as $L\to\infty$. Given that, our general line of arguments is similar to that of Ref.~\cite{BKE} devoted to noninteracting fermions in the $d$-dimensional simple cubic lattice $\mathbb Z^{d}$ with emphasis on the case $d=1$.

Our main results may be summarised as follows. For the (spinless) free Fermi gas in $\mathbb R^d$ at any $T>0$  we find the following two asymptotic large-scale expansions: the \emph{local thermal entropy} satisfies
\begin{equation}\label{S asympt}
S(T,L\Omega) = s(T) |\Omega| L^d + \eta(T,\partial\Omega) L^{d-1} + \ldots
\end{equation}
and the \emph{thermal} EE satisfies 
\begin{equation}\label{H asympt}
H(T,L\Omega) = 2\eta(T,\partial\Omega) L^{d-1} + \ldots\,,
\end{equation}
up to terms growing slower than $L^{d-1}$ as $L\to\infty$. Here, the bounded subregion $\Omega\subset \mathbb R^d$ may be rather general except that its boundary surface $\partial\Omega$ (if $d\ge2$) should be sufficiently smooth. For further assumptions see our theorem in Sec.~3 below. There we also make the definitions of the entropies $S(T,\Omega)$ and $H(T,\Omega)$ more precise and express the coefficient $\eta(T,\partial\Omega)$ in terms of a multiple integral. Nevertheless, in this Letter we concentrate on the physical aspects and publish the somewhat lengthy mathematical details in separate papers \cite{Sob2016, LSS3, Sob2015}. In the next section we just identify $s(T)$ and offer some explanations and comments.

\section{Physical meanings of the asymptotic coefficients and their dependence on temperature.} Not surprisingly, the leading asymptotic coefficient $s(T) |\Omega|$ in \eqref{S asympt} is nothing but the thermal entropy contained in $\Omega$ and $s(T)\geq 0$ is the (infinite-volume) \emph{thermal entropy density} or mean entropy. The latter is given by the thermodynamic relation
\begin{equation}\label{equ:s}
s(T) = \frac{\partial}{\partial T}\mathsf{p}(T),\quad \mathsf{p}(T):= \int_{\mathbb R} \text{d}E \, \mathcal N(E) f_T(E-\mu)\,,
\end{equation}
where the integral is the pressure of the free Fermi gas as a function of $T$ (and $\mu$). The quantity $\mathcal N(E) := (2\pi\hbar)^{-d} \int_{\mathbb R^d} \text{d}p\, \Theta(E-\varepsilon(p))$ defines the integrated density of states  $\mathcal N:\mathbb R\to [0,\infty[$ of the energy-mo\-men\-tum dispersion relation $\varepsilon:\mathbb R^d\to[0,\infty[$ which characterises the translation-invariant one-particle Hamiltonian of the free Fermi gas.\footnote{For convenience, we have assumed $\varepsilon(p)\ge0$ for all $p\in \mathbb R^d$ so that $\mathcal N(E) = 0$ if $E<0$. This is no loss of generality as long as $\varepsilon$ is bounded from below.} Here, $2\pi\hbar$ is Planck's constant and $\Theta$ is Heavi\-side's unit-step function. The second factor of the integrand in \eqref{equ:s} involves the Fermi function, $f_T:\mathbb R\to\,[0,1]$,  $f_T(E):=[1+\exp(E/T)]^{-1}$,  where from now on we put Boltzmann's constant $k_B=1$. From \eqref{equ:s} and Sommerfeld's asymptotic low-temperature expansion, 
\begin{equation}\label{Sommerfeld}
f_T(E) = \Theta(-E) -(\pi^2/6)\, T^2 \Theta''(E) +\ldots\
\end{equation}
(in distributional sense), we get the well-known formula
\begin{equation}\label{equ:s low T}
s(T) = (\pi^2/3)\, \mathcal N'(\mu) T + \ldots
\end{equation}
up to terms vanishing faster than $T$ as $T\downarrow0$. Eq.~\eqref{equ:s low T} holds, as it stands, at fixed chemical potential $\mu\in\mathbb R$. If  instead of $\mu$ the particle density $\rho>0$ is kept fixed, one has to invert the thermodynamic relation $\rho=\partial \mathsf{p}/\partial\mu$ between $\rho$ and $\mu$. By \eqref{Sommerfeld}, one thus finds at low temperatures another well-known formula 
\begin{equation}\label{mu at low T}
\mu(T,\rho) = \varepsilon_F - (\pi^2/6)\big[\mathcal N''(\varepsilon_F)/\mathcal N'(\varepsilon_F)\big] T^2 +\ldots\,,
\end{equation}
where $\varepsilon_F := \lim_{T\downarrow0}\mu(T,\rho)>0$ is the Fermi energy which satisfies $\rho = \mathcal N(\varepsilon_F)$. Consequently, Eq.~\eqref{equ:s low T} implies that $s(T) = (\pi^2/3) \mathcal N'(\varepsilon_F) T + \ldots$ for fixed  $\rho>0$. For the {\emph{ideal}} (non-relativistic, free) Fermi gas \cite{F26}, corresponding to the prime example $\varepsilon(p) = p^2/(2{\mathsf m})$ with ${\mathsf m}>0$ being the mass of each particle, we recall the explicit formula
\begin{equation}\label{IDS}
\mathcal N(E) = \Theta(E)\,\big[{\mathsf m}E/(2\pi\hbar^2)\big]^{d/2}/(d/2)!
\end{equation}
for its integrated density of states. Quantum effects dominate at low temperatures and become weaker at higher temperatures. Accordingly, the properties of the ideal Fermi gas approach those of the ideal Maxwell--Boltzmann gas in the high-temperature limit, $T\to\infty$. For example, in this limit the thermal entropy density of the ideal Fermi gas grows, to leading order, proportional to $T^{d/2}$ at fixed $\mu$ and proportional to $\ln(T)$ for fixed $\rho$, in symbols,
\begin{subequations}\label{ht1}
\begin{align}\label{ht1a}
	s(T) &\sim (T/T_{0})^{d/2}&& \text{($\mu$ fixed)}\,,\\
	s(T) &\sim \ln(T/T_0)&& \text{($\rho$ fixed)}.\label{ht1b}
\end{align}
\end{subequations}
Here, the constant $T_{0}>0$ is an arbitrary comparison temperature.

Returning to the expansions \eqref{S asympt} and \eqref{H asympt}, we note that 
the other asymptotic coefficient, $\eta(T,\partial\Omega)$, is also positive. 
On the one hand, it represents the thermal entropy on the 
boundary surface $\partial\Omega$ and determines the first-order 
 finite-size correction to the infinite-volume entropy (density). On the other hand, 
 $\eta(T,\partial\Omega)L^{d-1}$ 
 is half the thermal  EE to leading order in $L$. 
 Consequently, Eq.~\eqref{S asympt} shows that the 
 local entropy at $T>0$ displays a leading large-scale behaviour in 
 agreement with a ``volume law'' as it should be. 
 In contrast, Eqs.~\eqref{S asympt} and 
 \eqref{H asympt} show that the EE at 
 $T>0$ obeys a strict ``area law'': although 
 the two subregions $L\Omega$ and its complement $\mathbb R^d\setminus L\Omega$ carry (contrary to the case $T=0$) extremely different local entropies (namely $S(T,L\Omega)<\infty$ and $S(T,\mathbb R^d\setminus L\Omega)=\infty$), on their common boundary $\partial(L\Omega)$, or interface, the entropies are equal and proportional to $L^{d-1}$ as $L\to\infty$. Roughly phrased, the logarithmic enhancement $\ln(L)$ present \cite{LSS1,GiKl} in the large-scale behaviour of the ground-state EE, see \eqref{1} and \eqref{0EE} below, disappears when the temperature is raised from $T=0$ to $T>0$ because the Fermi surface ``grows soft''. Incidentally, we note that $\eta(T,\partial\Omega)$ does not depend on the choice of a condition imposed on the domain of the (quantum) Hamiltonian at the boundary $\partial\Omega$, because we work from the outset in the infinitely extended position space $\mathbb R^d$ and view all operators to act \emph{self-adjointly} on the associated one-particle Hilbert space $\text{L}^2(\mathbb R^d)$ of square-integrable functions $\psi:\mathbb R^d\to\mathbb C, q\mapsto \psi(q)$. 

The coefficient $\eta(T,\partial\Omega)$  is given by a rather complicated integral (see (\ref{u_alpha}--\ref{def:eta}) below), which, fortunately, is well-known in the theory of semiclassical expansions for traces of certain (truncated) Wiener--Hopf type operators, see Refs.~\cite{Wid80,Wid82,WidomLNM,R}. Interestingly, in the limit $T\downarrow0$  the coefficient $\eta(T,\partial\Omega)$ simplifies, displays a logarithmic singularity and takes (at fixed $\mu>0$ \footnote{If $\mu<0$, then $\eta(T,\partial\Omega)$ vanishes as $T\downarrow0$.}) the rather explicit form 
\begin{equation}\label{eta expansion}
\eta(T,\partial\Omega) = (1/12)\, J(\partial\Gamma_{\!\mu},\partial\Omega) \, \ln(T_{0}/T) + \ldots
\end{equation}
up to terms remaining bounded as $T\downarrow0$. Here, the level set $\partial\Gamma_{\!\mu} := \{p\in\mathbb R^d : \varepsilon(p)=\mu\}$ in momentum space is the (effective) Fermi surface corresponding to $\mu$. The factor $J(\partial\Gamma_{\!\mu},\partial\Omega)$ is defined as in Ref.~\cite{LSS1} and for $d\ge2$ given by the twofold surface integral
\begin{equation}\label{double surface}
J(\partial\Gamma_{\!\mu},\partial\Omega) := (2\pi\hbar)^{1-d} \int_{\partial \Gamma_{\!\mu}\times\partial \Omega}\text{d}\sigma({p}) \text{d}\tau({q}) \, \big|m(p)\cdot n(q)\big|\,.
\end{equation}
The vectors $m(p), n(q)\in\mathbb R^d$ denote the exterior unit normals at the points $p\in\partial\Gamma_{\!\mu}$ and $q\in\partial\Omega$, respectively. The canonical $(d-1)$-dimensional area measures on the surfaces $\partial\Gamma_{\!\mu}$ and $\partial\Omega$ are denoted by $\sigma$ and $\tau$, respectively. If we fix the particle density $\rho>0$ and use \eqref{mu at low T} in \eqref{eta expansion}, we arrive at \eqref{eta expansion} with $\mu$ replaced by $\varepsilon_F$. 

By identifying the large ratio ${T_{0}/T}$  inside the logarithm in \eqref{eta expansion} with the scaling parameter $L$, Eqs.~\eqref{S asympt} and \eqref{equ:s low T} give
\begin{equation}\label{0EE}
S(0,L\Omega) = (1/12)\, J(\partial\Gamma_{\!\mu},\partial\Omega) \, L^{d-1} \ln(L) + \ldots
\end{equation}
in agreement with the result for $T=0$ in Refs.~\cite{LSS1,GiKl} (resp. the corresponding expression with $\mu$ replaced by $\varepsilon_F$). For an isotropic dispersion relation $\varepsilon$ we know from Ref.~\cite{LSS1} that $J(\partial\Gamma_{\!\mu},\partial\Omega)$ is proportional to the area $|\partial\Omega|$.\footnote{In particular, for the ideal Fermi gas in $\mathbb R^d$ 
one  simply has $J(\partial\Gamma_{\!\mu},\partial\Omega)=2\,\mathcal{N}_{d-1}(\mu)|\partial\Omega|$, where $\mathcal{N}_{d-1}(E)$ is given by the right-hand side of \eqref{IDS} with $d$ replaced by $d-1$.} This is even true for $\eta(T,\partial\Omega)$ itself, at arbitrary  $T>0$. However, the emerging prefactor, the thermal entropy surface density, remains to be given by a multiple integral, see the remarks below the subsequent Eq.~\eqref{def:eta}.

As for the entropy density $s(T)$, the leading high-temperature behaviour of the coefficient $\eta(T,\partial\Omega)$ of the ideal Fermi gas depends on whether  $\mu$ or $\rho$ is kept fixed. More precisely, it follows from \eqref{def:eta} that
\begin{subequations}\label{ht2}
\begin{align}\label{ht2a}
	\eta(T,\partial\Omega)& \sim (T/T_{0})^{(d-1)/2}&& \text{($\mu$ fixed)},\\
	\eta(T,\partial\Omega)& \sim (T_{0}/T)^{1/2}&& \text{($\rho$ fixed)},\label{ht2b}
\end{align} 
\end{subequations}
as $T\to\infty$. Eq.~\eqref{ht2b} reflects the fact that the particles become effectively uncorrelated for fixed particle density at sufficiently high temperature.

\section{Precise definitions and formulations of results.} In order to define the local thermal entropy and the thermal EE precisely we first recall that the infinite-volume equilibrium state of the free Fermi gas at temperature $T>0$ and chemical potential $\mu\in\mathbb R$ is quasi-Gaussian (in other words, quasi-free) and uniquely determined by its reduced one-particle density operator $f_T(\varepsilon(P)-\mu\mathbbm 1)$ on $\text{L}^2(\mathbb R^d)$. Here, $P:= -i\hbar\partial/\partial q$ is the canonical-momentum operator, $\varepsilon(P)\ge0$ the one-particle quantum Hamiltonian and $\mathbbm 1$ the identity operator. The local (or truncated) version 
\begin{equation}\label{15}
D(f_T,\Omega) := \mathbbm 1_\Omega\,f_T(\varepsilon(P)-\mu\mathbbm 1)\,\mathbbm 1_\Omega
\end{equation}
of the density operator then characterises the quasi-Gaussian substate obtained from the equilibrium state by spatial reduction to $\Omega\subseteq\mathbb R^d$. Here, $\mathbbm 1_\Omega : \text{L}^2(\mathbb R^d)\to\text{L}^2(\mathbb R^d)$ denotes the  projection operator associated with the indicator function of $\Omega$, that is, $(\mathbbm 1_\Omega \psi)(q) := \psi(q)$ if $q\in\Omega$ and $0$ otherwise for all $\psi\in \text{L}^2(\mathbb R^d)$. 
Next we recall the \emph{binary entropy function} $h: [0,1] \to[0,\ln(2)]$ defined by 
\begin{equation}\label{vonneumann:eq}
h(0):=h(1):=0\ \ \textup{and}\ \ h(t) :=-t\ln{(t)} - (1-t) \ln{(1-t)} \ \ \textup{if}\ \ t\in\,]0,1[ .
\end{equation}
The positivity 
\begin{equation}\label{htilde:eq}
\widetilde{h}(\lambda,r,t) := h\big((1-\lambda)r 
+ \lambda t\big) - (1-\lambda)h(r) -\lambda h(t) \ge0
\end{equation}
for all  $\lambda, r,t \in\,[0,1]$ is equivalent to the concavity of $h$. In fact, $h$ is even operator-concave 
\cite{Ando}. The \emph {local thermal} (von Neumann) \emph{entropy} is now given as the trace
\begin{equation}\label{s_alpha}
S(T,\Omega) := \mathrm{Tr}[ \mathbbm 1_\Omega \, 
h(D(f_T,\Omega))\mathbbm 1_\Omega] = \mathrm{Tr}[ \,h(D(f_T,\Omega))],
\end{equation}
see, for example, \cite{HLS}. The second equality in \eqref{s_alpha} follows from $h(0)=0$.

The definition of the thermal (von Neumann) EE requires two steps. In the first step we introduce the ``entropic'' operator difference 
\begin{equation} \label{def:Delta}
\Delta(T,\Omega) := \mathbbm 1_\Omega h(D(f_T,\Omega))\mathbbm 1_\Omega - D(h\circ f_T,\Omega) \,.
\end{equation}
Here, the operator $D(h\circ f_T,\Omega)$ is obtained from \eqref{15} by replacing the Fermi function $f_T$ with the composed function $h\circ f_T$ defined by $(h\circ f_T)(E) := h(f_T(E))$ for all $E\in\mathbb R$. From the operator concavity of the function $h$ and from Refs.~\cite{Ando, Davis} we get the operator positivity $ \Delta(T,\Omega) \geq 0$. In the second step we define the \emph{thermal} EE for a \emph{bounded} $\Omega\subset\mathbb R^d$ as the sum of two positive traces
\begin{equation} \label{def:EE}
H(T,\Omega) := \mathrm{Tr}\,\Delta(T,\Omega) + \mathrm{Tr}\,\Delta(T,\mathbb R^d\setminus\Omega)\,.
\end{equation}
This is the precise version of  \eqref{2}. Arguments as in Ref.~\cite{Sob2015b} show that even the second trace is finite, although the (positive) minuend and the (positive) subtrahend of $\Delta(T,\mathbb R^d\setminus\Omega)$ have both an infinite trace. Therefore, we arrive at the (in)equalities
\begin{equation} \label{chain}
0\le H(T,\Omega) = H(T,\mathbb R^d\setminus\Omega) <\infty \, .
\end{equation}
Moreover, in the limit $T\downarrow0$ we get back to Eq. \eqref{1} by observing that $h\circ f_0=0$, confer \eqref{Sommerfeld}.

To explain the coefficient $\eta(T,\partial\Omega)$ of the subleading asymptotic behaviour of $S(T,L\Omega)$ and of the leading behaviour of $H(T,L\Omega)$ as $L\to\infty$, we need some auxiliary definitions. In contrast to the well-known leading ``volume term''  in \eqref{S asympt} the subleading ``area term'' is new and rather complicated. It cannot be obtained from simple heuristic considerations, not even for $d=1$. But it can be derived from the semiclassical ``area coefficient'' in Refs.~\cite{Wid80,Wid82,WidomLNM,R} by observing that the traces in \eqref{s_alpha} and \eqref{def:EE}, with $\Omega$ replaced by $L\Omega$, depend on Planck's constant $2\pi\hbar$ and the scaling parameter $L$ only via the ratio $\hbar/L$ which can be seen by a (unitary) dilatation. In order to recall this ``area coefficient''  from the mentioned literature we first define the function  $U: [0,1]\times[0,1]\to[0,\infty[$, $(r,t)\mapsto U(r,t)$ by 

\begin{equation} \label{u_alpha}
U(r,t):=\frac{1}{8\pi^2}\,\int_0^1 \text{d}\lambda\, \frac{\widetilde{h}(\lambda,r,t)}{\lambda(1-\lambda)} \,,
\end{equation}
see \eqref{vonneumann:eq} and \eqref{htilde:eq} for the definition of $\widetilde h$. 
Then we consider the integral
\begin{equation}\label{def:B}
\mathcal U[g] := \underset{\delta\downarrow 0}\lim\int_{\mathbb R\times\mathbb  R} \text{d}v \,\text{d}w \,\Theta ( |v-w|-\delta ) \;\frac{U\big(g(v),g(w)\big)}{(v-w)^2} \,
\end{equation}
defined, in the principal-value sense, for smooth functions $g:\mathbb R\to\mathbb [0,1]$. We observe that 
$\mathcal U[g]\geq 0$ since $U(r,t)\geq 0$ due to the concavity of $h$. 
If the function $h$ were smooth, then $\mathcal U[g]$ could be defined
as a standard Riemannian integral, whose finiteness could be easily checked (for smooth $g$).   
For the function \eqref{vonneumann:eq} however, as well as for certain 
other functions (such as $t\mapsto t^\alpha(1-t)^\alpha$,\ 
$\alpha\in\,]0, 1[$ ) being continuous but not differentiable  
at the points $t = 0$ and $t=1$, the finiteness of $\mathcal U[g]$ is a non-trivial matter. 
This and other relevant properties of the integral \eqref{def:B} are 
investigated in \cite{Sob2016}. 

If $d\ge2$, then we consider for $g$ at given $T>0$ the $2(d-1)$-parameter family of functions 
$f_{T;(p,q)}:\mathbb R\to\,[0,1]$ defined in terms of the Fermi function by
$f_{T;(p,q)}(v) := f_T\big(\varepsilon(p+v\,n(q))-\mu\big)$, where $v\in\mathbb R, q\in\partial\Omega$ and 
$p\in \mathsf{T}_q^*(\partial\Omega)$ with $\mathsf{T}_q^*(\partial\Omega)\cong \mathbb R^{d-1}$ 
being the dual space of the $(d-1)$-dimensional tangent space of $\partial\Omega$ at the point $q$. 
The vector $n(q)\in\mathbb R^{d}$ and the subsequent area measure $\tau$ have the same meanings as in \eqref{double surface}. 
Finally, we define
\begin{equation}\label{def:eta}
\eta(T,\partial\Omega):=(2\pi\hbar)^{1-d}\\
\int_{\partial\Omega}\text{d}\tau(q) \\
\int_{\mathsf{T}_q^*(\partial\Omega)} \text{d}p\; \mathcal U[f_{T;(p,q)}] \,.
\end{equation} 
If the dispersion relation $\varepsilon$ is isotropic, then the functions $f_{T;(p,q)}$ 
do not depend on the parameter $q\in\partial\Omega$ due to the 
orthogonality of $p\in \mathsf{T}^*_{q}(\partial\Omega)$ and $n(q)$. 
Consequently, the surface area $|\partial\Omega|$ 
can be factored out on the right-hand side of \eqref{def:eta}. 
Nevertheless, the multiple integral underlying 
$\eta(T,\partial \Omega)$ remains to be a fourfold one for $d\geq2$. 
For $d=1$ the set $\Omega\subset\mathbb R$ is the union of finitely many pairwise disjoint bounded intervals, according to an assumption in the subsequent theorem. 
Then $|\partial\Omega|$ equals the (even) number of all endpoints of the constituent intervals and one has 
$\eta(T,\partial\Omega) = \mathcal U[f_T\circ(\varepsilon-\mu)]\, |\partial\Omega|$, 
which still involves a threefold integral.

Now we are prepared to state our central ``technical'' result as the following

\begin{thm}
For any temperature $T>0$, for any bounded subregion $\Omega\subset\mathbb R^d$ with finitely many connected components and (if $d\ge2$) piecewise smooth boundary surface $\partial\Omega$ (see Ref.~\cite{Sob3}) and for any smooth and polynomially bounded dispersion relation $\varepsilon$ the trace of $\Delta(T,\Lambda)$ is finite and positive 
for $\Lambda = \Omega$ or $\Lambda = \mathbb R^d\setminus\Omega$ with the same leading term in both large-scale expansions
\begin{equation}\label{expansion}\mathrm{Tr}\, \Delta(T,L\Lambda) = \eta(T,\partial\Omega) \,L^{d-1} + \ldots\,,
\end{equation}
up to terms growing slower than $L^{d-1}$ as $L\to\infty$.
\end{thm}

Our proof goes as follows. At first we ``smooth out" the function $h$, which enables us to refer for the formula \eqref{expansion} directly to available literature, notably to \cite{Wid80} or \cite{WidomLNM} (see, in particular, Chap.~I and V in \cite{WidomLNM}). To return to the original non-smooth function $h$, we then ``close" the asymptotics using the estimates obtained in \cite{Sob2015b} for non-smooth functions of operators of the type $D(f, \Omega)$. This strategy is similar to the one applied in \cite{LSS1}. Full mathematical details will be published elsewhere \cite{LSS3, Sob2015}. There we also establish the validity of formula \eqref{expansion} even if $L\to\infty$ and $T\downarrow0$ \emph{simultaneously} provided that $L T/T_{0} \ge 1$ throughout. In our view, these observations provide a deeper insight into the low-temperature behaviour of the thermal EE. 

Combining \eqref{def:EE} and \eqref{expansion} immediately gives the claimed large-scale behaviour  \eqref{H asympt} of the thermal EE. From \eqref{expansion} we also infer the claimed two-term large-scale behaviour  \eqref{S asympt} of the local  thermal entropy \eqref{s_alpha} by observing
\begin{align}
S(T,L\Omega)&=\mathrm{Tr} \,D(h\circ f_T,L\Omega) + \mathrm{Tr}\,\Delta(T,L\Omega)\notag
\\
&=s(T) |\Omega| L^d+\mathrm{Tr}\,\Delta(T,L\Omega)\notag
\\ \label{27}&=s(T)\, |\Omega|\, L^d + \eta(T,\partial\Omega) \,L^{d-1} + \ldots\,,
\end{align}
where we have used the identities 
\begin{equation}
(2\pi\hbar)^{-d} \int_{\mathbb R^d} \text{d}p\, h\big(f_T(\varepsilon(p)-\mu)\big) \\ = 
\int_{\mathbb R} \text{d}E\,{\mathcal N}'(E) h\big(f_T(E-\mu)\big) = s(T)
\end{equation}
in the second equality. The leading large-scale behaviour of the local entropy $S(T,L\Omega)$ at $T>0$ was first proved in \cite{PS,BHK}. The subleading correction of the order $L^{d-1}$ in \eqref{27} is new.

\section{Summary, discussion and an open problem.}

For the free Fermi gas in multidimensional continuous space $\mathbb R^d$  in thermal equilibrium at temperature $T>0$  and for a given bounded subregion $\Omega\subset\mathbb R^d$ we carefully distinguish between the \emph {local thermal entropy} $S(T,\Omega)\ge s(T)|\Omega|$ and the \emph{thermal spatially bipartite entanglement entropy} $H(T,\Omega)\ge 0$. Their large-scale behaviours  \eqref{S asympt} and \eqref{H asympt} contain the thermal entropy density  $s(T)\ge 0$ and the new asymptotic coefficient  $\eta(T,\partial\Omega)\ge 0$ as two characteristics of the rather old free Fermi-gas model \cite{F26}. The results \eqref{S asympt} and \eqref{H asympt} are physically even more relevant than the corresponding
ground-state results \eqref{0EE} and \eqref{1}, because in real gases the temperature is never strictly zero. Furthermore, the new results \eqref{S asympt} and \eqref{H asympt} deepen our understanding of the older ones \eqref{0EE} and \eqref{1} by observing that $s(0)=0$ and that $\eta(T, \partial\Omega)$ 
diverges logarithmically as $T\downarrow0$ according to \eqref{eta expansion}. 
A result similar to \eqref{eta expansion} for noninteracting fermions in the one-dimensional lattice $\mathbb Z^{1}$  was derived in \cite{BKE}, but without an explicit prefactor. As in  \cite{LSS1,BKE}, many of our present results extend to the whole one-parameter family of (quantum) R\'enyi entropies, see \cite{LSS3, Sob2015}.
 
Finally, we check whether the exact asymptotic results \eqref{0EE}, \eqref{S asympt}, \eqref{equ:s low T} and \eqref{eta expansion} for the local thermal entropy $S(T,L\Omega)$ can be descibed consistently by a so-called universal crossover formula as obtained in \cite{K, Swin3,SwinSent,CC1,CC2,CCT2} by arguments from a suitable conformal field theory, at least for $d=1$. Such a formula has the appealing form
\begin{equation}\label{korepin}
S(T,L\Omega)= L^{d-1} A \ln\Big[ \frac{T_{0}}{T}\sinh \Big(L \frac{T}{T_{0}}\Big)\Big]
\end{equation}
with suitable constants $A\geq 0$ and $T_{0}>0$ not depending on $L$ and $T$.
Remarkably, if we choose $A=(1/12)\,J(\partial\Gamma_{\!\mu},\partial\Omega)$ and
$T_{0} = 3A/[\pi^{2}\mathcal{N}'(\mu)|\Omega|]$, then we find consistency, provided that $L\gg 1$  and $T \ll T_{0}$. We conclude that \eqref{korepin} correctly reflects asymptotic properties of the free Fermi gas in $\mathbb R^{d}$ if $L$ is large and $T$ is small, but does, for example, not reproduce the large-$T$ behaviours \eqref{ht1} and \eqref{ht2} (of the ideal Fermi gas). However, Eq. \eqref{korepin} suggests the scaling-type formula
\begin{equation}\label{scaling}
\lim\limits_{L\to\infty}\frac{1}{L^{d-1}}\Big[S\Big(\frac{xT_{0}}{L},L\Omega\Big)-S(0,L\Omega)\Big] = A \ln\Big[\frac{\sinh(x)}{x}\Big]
\end{equation}
for \emph{any} $x>0$. Although we have a result for the simultaneous limits $L\to\infty$ and $T\downarrow0$ with $x=LT/T_0 \ (>1)$ kept fixed, see a remark below the above theorem, at present we do not know whether \eqref{scaling} or a similar formula follows from the ``microscopic'' definition \eqref{s_alpha}, not even for $d=1$.
 
\ack We thank Ingo Peschel (FU Ber\-lin) for valuable discussions.

\section*{References}

\bibliography{biblio2}

\begin{thebibliography}{10}

\bibitem{HHHH}
R.~Horodecki, P.~Horodecki, M.~Horodecki, and K.~Horo\-decki.
\newblock Quantum entanglement.
\newblock {\em Rev. Mod. Phys.}, 81:865--942, 2009.

\bibitem{Sach}
S.~Sachdev.
\newblock {\em {Quantum Phase Transitions, 2nd edition}}.
\newblock Cambridge University Press, Cambridge, United Kingdom, 2011.

\bibitem{Vidal}
G.~Vidal, J.I. Latorre, E.~Rico, and A.~Kitaev.
\newblock Entanglement in quantum critical phenomena.
\newblock {\em Phys. Rev. Lett.}, 90:227902, 2003.

\bibitem{Chak}
S.~Chakravarty.
\newblock {Scaling of von Neumann entropy at the Anderson transition}.
\newblock {\em Int. J. Mod. Phys. B}, 24:1823--1840, 2010.

\bibitem{Pastur}
L.~Pastur and V.~Slavin.
\newblock {Area law scaling for the entropy of disordered quasifree fermions}.
\newblock {\em Phys. Rev. Lett.}, 113:150404, 2014.

\bibitem{ECP}
J.~Eisert, M.~Cramer, and M.B. Plenio.
\newblock Area laws for the entanglement entropy -- a review.
\newblock {\em Rev. Mod. Phys.}, 82:277--306, 2010.

\bibitem{H}
M.B. Hastings.
\newblock {Entropy and entanglement in quantum ground states}.
\newblock {\em Phys. Rev. B}, 76:035114, 2007.

\bibitem{LSS1}
H.~Leschke, A.V. Sobolev, and W.~Spitzer.
\newblock {Scaling of R\'enyi entanglement entropies of the free {Fermi}-gas
  ground state: a rigorous proof}.
\newblock {\em Phys. Rev. Lett.}, 112:160403, 2014.

\bibitem{GiKl}
D.~Gioev and I.~Klich.
\newblock {Entanglement entropy of fermions in any dimension and the {Widom}
  conjecture}.
\newblock {\em Phys. Rev. Lett.}, 96:100503, 2006.

\bibitem{BKE}
H.~Bernigau, M.J. Kastoryano, and J.~Eisert.
\newblock Mutual information area laws for thermal free fermions.
\newblock {\em J. Stat. Mech.: Theory and Experiment}, 2015:P02008, 2015.

\bibitem{Sob2016}
A.V. Sobolev.
\newblock On a coefficient in trace formulas for {Wiener--Hopf} operators.
\newblock {\em arXiv:1601.00463 [math.SP]}, 2016.

\bibitem{LSS3}
H.~Leschke, A.V. Sobolev, and W.~Spitzer.
\newblock {Trace formulas for {Wiener--Hopf} operators with applications to
  entropies of free fermionic equilibrium states}.
\newblock {\em arXiv: 1605.04429 [math.SP]}, 2016.

\bibitem{Sob2015}
A.V. Sobolev.
\newblock Trace formula for multidimensional {Wiener--Hopf} operators.
\newblock {\em {In preparation}}, 2016.

\bibitem{F26}
E.~Fermi.
\newblock Zur {Q}uantelung des idealen einatomigen {G}ases.
\newblock {\em Z. Physik}, 36:902--912, 1926.

\bibitem{Wid80}
H.~Widom.
\newblock {Szeg\H{o}}'s limit theorem: the higher-di\-men\-sional matrix case.
\newblock {\em J. Funct. Anal.}, 39:182--198, 1980.

\bibitem{Wid82}
H.~Widom.
\newblock A trace formula for {Wiener}--{Hopf} operators.
\newblock {\em J. Operator Theory: Adv. Appl.}, 8:279--298, 1982.

\bibitem{WidomLNM}
H.~Widom.
\newblock {\em {Asymptotic Expansions for Pseudodifferential Operators on
  Bounded Domains}}.
\newblock Springer--Verlag, Berlin, 1985.

\bibitem{R}
R.~Roccaforte.
\newblock {Asymptotic expansions of traces for certain convolution operators}.
\newblock {\em Trans. Amer. Math. Soc.}, 285:581--602, 1984.

\bibitem{Ando}
T.~Ando.
\newblock Concavity of certain maps on positive definite matrices and
  applications to {Hadamard} products.
\newblock {\em Lin. Alg. and Appl.}, 26:203--241, 1979.

\bibitem{HLS}
R.~Helling, H.~Leschke, and W.~Spitzer.
\newblock A special case of a conjecture by {Widom} with implications to
  fermionic entanglement entropy.
\newblock {\em Int. Math. Res. Notices}, 2011:1451--1482, 2011.

\bibitem{Davis}
C.~Davis.
\newblock A {Schwarz} inequality for convex operator functions.
\newblock {\em Proc. Amer. Math. Soc.}, 8:42--44, 1957.

\bibitem{Sob2015b}
A.V. Sobolev.
\newblock Functions of self-adjoint operators in ideals of compact operators.
\newblock {\em arXiv: 1504.07261 [math.SP]}, 2015.

\bibitem{Sob3}
A.V. Sobolev.
\newblock {Wiener--Hopf} operators in higher dimensions: the {Widom} conjecture
  for piece-wise smooth domains.
\newblock {\em Integr. Equ. Oper. Theory}, 81:435--449, 2015.

\bibitem{PS}
Y.M. Park and H.H. Shin.
\newblock {Dynamical entropy of space translations of {CAR} and {CCR} algebras
  with respect to quasi-free states}.
\newblock {\em Commun. Math. Phys.}, 152:497--537, 1993.

\bibitem{BHK}
F.~Benatti, T.~Hudetz, and A.~Knauf.
\newblock Quantum chaos and dynamical entropy.
\newblock {\em Commun. Math. Phys.}, 198:607--688, 1998.

\bibitem{K}
V.E. Korepin.
\newblock {Universality of entropy scaling in one dimensional gapless models}.
\newblock {\em Phys. Rev. Lett.}, 92:096402, 2004.

\bibitem{Swin3}
B.~Swingle.
\newblock Conformal field theory approach to {Fermi} liquids and other highly
  entangled states.
\newblock {\em Phys. Rev. B}, 86:035116, 2012.

\bibitem{SwinSent}
B.~Swingle and T.~Senthil.
\newblock Universal crossover between entanglement entropy and thermal entropy.
\newblock {\em Phys. Rev. B}, 87:045123, 2013.

\bibitem{CC1}
P.~Calabrese and J.~Cardy.
\newblock Entanglement entropy and quantum field theory.
\newblock {\em J. Stat. Mech.: Theory and Experiment}, 2004:P06002, 2004.

\bibitem{CC2}
P.~Calabrese and J.~Cardy.
\newblock Entanglement entropy and conformal field theory.
\newblock {\em J. Phys. A: Math. Theor.}, 42:504005, 2009.

\bibitem{CCT2}
P.~Calabrese, J.~Cardy, and E.~Tonni.
\newblock Finite temperature entanglement negativity in conformal field theory.
\newblock {\em J. Phys. A: Math. Theor.}, 48:015006, 2015.

\end{thebibliography}

\end{document}